# Beyond Companionship: Robotic Pets as Embodied Communication Media for Older Adults

A Formative Interview Study of Remote Presence and Relationship Maintenance

Delong Du
delong.du@uni-siegen.de
University of Siegen
Siegen, Germany

Sara Gilda Amirhajlou
sara.amirhajlou@student.uni-siegen.de
University of Siegen
Siegen, Germany

Akwasi Gyabaah
akwasi.gyabaah@student.uni-siegen.de
University of Siegen
Siegen, Germany

Richard Paluch
richard.paluch@uni-siegen.de
University of Siegen
Siegen, Germany

Claudia Müller
claudia.mueller@uni-siegen.de
University of Siegen
Siegen, Germany

## ABSTRACT

Robotic pets for older adults are typically studied as companions, with the older person positioned as the robot's primary interaction partner. This framing overlooks another role: a petlike robot can mediate relationships between people. We report a formative qualitative interview study with six adults aged 63-77 in Germany. Interviews examined participants' existing communication practices, experiences of being alone, and responses to scenarios in which a robotic pet could carry voice messages, support calls, move through the home, or convey a remote family member's actions. Thematic qualitative content analysis identified four findings. Participants maintained relationships through flexible combinations of telephone, messaging, and video calls; imagined petlike movement, proximity, voice, and touch-like gestures adding embodied presence to those channels; combined expectations of communication, companionship, and practical assistance; and made acceptance conditional on customization, hygiene, ease of use, and human control. Drawing together HCI research on mediated intimacy, CSCW research on awareness and care networks, and HRI research on social and telepresence robots, we contribute (1) an empirical account of how older adults imagine petlike embodiment carrying another person's presence, (2) the concept of robotic pets as embodied communication media, and (3) a four-part sensitizing framework organized around relational purpose, source of agency, mode of embodiment, and governance. The study identifies design hypotheses; it does not test a robot or establish effects on loneliness, depression, or relationship quality.

## 1 INTRODUCTION

Social relationships remain important throughout later life. Social disconnectedness and perceived isolation are associated with poorer health and well-being among older adults, although the form and strength of these associations vary across people and circumstances [9, 15, 33]. Digital communication can help people sustain existing ties across distance, but a telephone or video call does not provide physical co-presence, movement through a shared environment, or touch. These missing qualities matter when family members and friends want to express affection or take part in everyday routines rather than only exchange information.

Robotic pets are usually discussed as companions for an older person. Studies of PARO, Joy for All pets, and other zoomorphic robots have examined emotional attachment, engagement, acceptability, and possible effects on loneliness or depression [12, 17, 18, 20]. This work has established that petlike form and behavior can support approachable, tactile interaction. It has also shown that use is social: relatives, care workers, and other residents interpret, introduce, share, and sometimes animate the robot [2, 27, 30]. A robotic pet therefore need not be the endpoint of a relationship. It can become a medium through which people act toward one another.

Telepresence research offers a complementary perspective. Mobile telepresence robots allow a remote person to speak, see, and move within another location. Studies with older adults indicate that this mobility can increase perceived presence and support family contact, while also raising concerns about privacy, asymmetrical control, and intrusion [4, 8, 23, 25]. Most telepresence platforms, however, use a screen-on-wheels form. They are designed to transmit a remote person's audiovisual presence rather than to draw on the familiarity, tactility, and affective expressiveness of a pet.

This paper examines the overlap between these two design spaces. We ask:

*How might a robotic pet mediate communication and a sense of presence between older adults and people they care about?*

We address this question through semi-structured interviews with six older adults in Germany. Participants discussed their existing relationship-maintenance practices and imagined how a robotic pet might support communication, companionship, and daily life. The paper makes three contributions. First, it provides an empirical account of how participants connected petlike embodiment with remote family presence, especially through voice, movement, proximity, and touch-like action. Second, it develops the concept of the robotic pet as embodied communication media: an object that may be experienced at different moments as a pet, a device, and an expression of another person. Third, it offers a four-part sensitizing framework—relational purpose, source of agency, mode of embodiment, and governance—for comparing designs and structuring future evaluations. Together, these contributions shift the research question from whether an older adult relates to a robot toward how people may relate to one another through a robot.

Because the study was formative and did not involve a working robot, its claims are deliberately limited. The findings describe expectations and design possibilities among a small, technically interested sample. They do not demonstrate that a robotic pet improves mental health, reduces loneliness, or strengthens relationships.

## 2 RELATED WORK ACROSS HCI CSCW AND HRI

### 2.1 HCI perspectives on relationship maintenance and mediated touch

Relationships can provide companionship, emotional support, practical help, and a sense of belonging. Social disconnectedness and perceived isolation are related but distinct: a person may have few contacts without feeling lonely, or feel lonely despite frequent contact [9]. Family ties also contain support, obligation, tension, and negotiation, so technology should not assume that more contact is always desirable or that family members share the same goals [29, 33].

HCI research treats relationship maintenance as an interaction-design problem rather than a simple matter of increasing contact. Vetere et al. [34] described intimacy as something technologies can mediate through subtle, relationship-specific exchanges, and Hassenzahl et al. [14] analyzed how interactive artifacts can make intimate experience tangible. Chien and Hassenzahl [6] further showed how communication technologies can support recurring practices of intimacy across distance, while Nakanishi et al. [24] demonstrated that adding mediated touch to video communication can increase social telepresence. Together, this work distinguishes information exchange from small embodied expressions of care—the distinction that motivates our focus on petlike movement, proximity, and touch-like action.

Older adults use combinations of telephone calls, messaging, video calls, and in-person visits to maintain relationships. The suitability of a medium depends on a person's abilities, habits, devices, relationship, and purpose for communicating. Technology can extend an existing relationship, but it can also introduce work: charging a device, learning an interface, scheduling a call, troubleshooting connectivity, or managing unwanted access. Designs for social connection therefore need to fit an ecology of existing practices rather than assume a blank slate.

### 2.2 HRI and robotics research on robotic companions

Robotic pets range from relatively simple animatronic cats and dogs to responsive systems such as PARO and AIBO. Their petlike form can invite stroking, holding, grooming, naming, and talk. These interactions differ from using a conventional computer because the robot occupies space, moves, and can be approached as a social object [10, 22]. HRI research also shows that sustained social-robot interaction depends on adaptation, continuity, and the fit between a robot's behavior and a person's changing expectations [21]. From an HCI perspective, autonomous and socially expressive systems can be experienced as counterparts rather than tools, a class of interaction described as otherware [13].

Research with older adults has reported positive experiences with robotic pets, including enjoyment, calming interaction, and opportunities for social engagement [5, 17, 18, 28]. The evidence is heterogeneous, however. Devices, settings, participants, exposure periods, and outcomes vary widely. A robotic pet may be welcomed as a toy or companion by one person and rejected as childish, deceptive, or irrelevant by another [7, 20, 27].

Robotic pets also shape interaction among people. In group settings, PARO has affected conversation and participation among residents and facilitators [30]. Some owners share a robotic pet or use it as a topic of conversation [17]. These observations move the analysis beyond a human-robot dyad and toward a network of relationships and practices.

### 2.3 CSCW perspectives on care networks and situated use

CSCW research cautions against treating a care technology as an isolated device used by one individual. Classic work on shared systems defined awareness of others' activities as a resource for coordinating interdependent action [11]. In a connected home, awareness also has a protective function: people need to know who is present through a device, what that person can do, and when an action is automated. Robots become usable through coordination among residents, relatives, care workers, researchers, and institutions. Ethnographic studies show that staff members introduce robots, interpret their behavior, manage access, repair breakdowns, and integrate them into care routines [2, 3, 27]. Hornecker et al. [16] describe this as interaction work: the functioning of care technology is inseparable from the verbal, manual, and emotional practices through which caregivers and residents enact it.

Communication technologies also redistribute work. Schwaninger et al. [31] found that video contact in care homes relied on a triad of residents, relatives, and care workers, with staff arranging calls, setting up devices, and often remaining present. For a remotely accessible robotic pet, these CSCW findings imply that privacy, availability, maintenance, and responsibility cannot be designed only at the interface level. They must be negotiated across a care network.

## 2.4 Telerobots and social mediator robots

Telepresence robots explicitly mediate interaction between people. Their value lies in allowing a remote partner to look around, move, join an activity, and appear in a local space. A feasibility study in residential care found that a telepresence robot could support family communication for people with dementia, while revealing technical and organizational challenges [23]. A year-long home study similarly described telepresence as a way to share experiences and extend family presence without replacing in-person contact [4]. Work in residential care has emphasized privacy, room access, mobility constraints, and the need for ethical guidance [25].

HCI research on telerobotics asks what embodied systems can add beyond reproducing a video call. Albers et al. [1] developed speculative telerobot scenarios around reciprocal, short, and activity-based forms of informal care, highlighting relational autonomy and the asymmetries within care relationships. This focus aligns with our participants' interest in low-intensity gestures and remote family presence through a petlike body.

Recent studies sharpen the relevance of agency. Conde et al. [8] found that older adults' attitudes toward robot-mediated communication differed substantially and that designs should fit existing skills, lifestyles, and social networks. Simão et al. [32] showed how older residents could use a robot to send messages and objects, with control over the robot supporting self-expression but also producing participation asymmetries. Noguchi et al. [26] found that the way a mediator robot conveys sensitive messages can affect older adults' anxiety about self-disclosure. Together, this work suggests that a robot's social value depends on who initiates an interaction, who controls the robot, how that control is displayed, and whether each person can accept, refuse, or reshape the encounter. Petlike robots may add a further layer to robot-mediated communication. A soft body, animal-like movement, and touchable surfaces could communicate affection or presence without requiring a remote person's face to remain on a screen. At the same time, petlike behavior could obscure whether an action comes from the remote partner, from automation, or from the device itself. This ambiguity motivates our focus on embodied communication, legibility, and reciprocal control.

# 3 STUDY DESIGN

## 3.1 Participants and recruitment

We conducted semi-structured interviews with six adults in Germany aged 63-77. Participants were recruited through an online forum that invited people interested in aging and technology. All six were described in the original study as having relevant prior experience and an interest in the topic. The sample comprised five men and one woman; five participants were married and one was single. Table 1 reproduces the reported demographic information. This recruitment strategy was appropriate for exploratory design work but shaped the findings. Participants were self-selected and technically interested. Most were men, and most were married.

**Table 1: Participant characteristics.**

| Participant | Age | Gender | Marital status |
|---|---|---|---|
| P1 | 77 | Man | Single |
| P2 | 75 | Woman | Married |
| P3 | 71 | Man | Married |
| P4 | 63 | Man | Married |
| P5 | 72 | Man | Married |
| P6 | 67 | Man | Married |

The study therefore does not represent older adults who have limited digital experience, live with cognitive impairment, reside in institutional care, or have different family and household arrangements.

## 3.2 Interview procedure

Interviews were conducted one-to-one over Zoom and lasted between 40 and 90 minutes. Participants consented to recording. The semi-structured guide moved from current experience to speculative design. It covered:

- everyday contact with family members and friends, including positive memories and preferred communication channels;
- experiences of being alone or waiting for a response from someone;
- initial reactions to the idea of a robotic pet and functions participants would want it to perform;
- design scenarios in which the pet could carry a voice message, support a call, move with the user, or be controlled by a family member; and
- demographic questions and debriefing.

The questions were intended to elicit design possibilities, not to evaluate a particular robot. Participants did not interact with a prototype. Their responses therefore concern imagined use and anticipated value.

## 3.3 Analysis

The recordings were transcribed and managed in MAXQDA. We used a thematic qualitative content analysis informed by Kuckartz [19]. Analysis combined deductive attention to the interview topics with inductive coding of participants' accounts and design suggestions. Initial codes included communication practices, positive and negative family interactions, coping when alone, desired companionship, preferred robot tasks, voice and movement, anticipated family responses, and practical design requirements. Related codes were compared and grouped into broader themes.

For this revision, we reorganized the reported material around the paper's central analytical distinction: a robotic pet can be a companion, a practical device, and a medium for another person's presence. The quotations reproduced below are the English-language excerpts reported in the original paper. The analysis remains bounded by those data; no new interviews or outcome measures were added.

# 4 FINDINGS

## 4.1 Relationships were maintained through a flexible media ecology

Participants described communication with family members as important whether relatives lived nearby or at a distance. They did not rely on a single channel. Instead, they selected among telephone calls, messaging, and video calls according to the situation and what they wanted to convey. P5 described this arrangement as integrated communication: "We can only communicate by phone … We use all possible communication channels depending on what we want to do or transport. We have integrated communication."

This account positions a robotic pet as one possible element in an existing media ecology. Participants did not describe family contact as absent or imagine the robot replacing their current devices. The design opportunity was to make some interactions easier or qualitatively different. This distinction matters because a new device must justify the additional effort it creates. It should either reduce interaction costs, such as simplifying voice messaging for a relative who does not text, or add a quality that the telephone and smartphone do not provide.

P3 made the accessibility issue concrete when discussing an older relative: "My mom who is much older can only call us [rather] than texting or video call." A pet that supports a familiar spoken interaction might lower the threshold for sending or receiving a message. Yet the comment also warns against treating age as a uniform condition. The participants themselves used several technologies and differed in what they considered convenient.

Participants consistently framed the robotic pet as an addition to human contact. They did not want it to become a justification for fewer calls or visits. Its value depended on connecting people or supporting daily life between encounters.

## 4.2 Petlike embodiment could carry a remote person's presence

The most distinctive design idea concerned telepresence without a conventional screen-on-wheels form. Participants imagined that a remote family member could speak through the pet, control its movement, or trigger a gentle gesture. These actions could make communication feel located in the room and associated with a particular person.

P2 connected remote control with affectionate touch: "I can feel my family gently touch me if they remotely control the pet to show me their love." P3 imagined voice and identity combining in the object: "I start to feel the pet can feel like my son because he can send me voice messages and basically talk to me through the dog." P5 made a similar association: "Oh, I am thinking this pet is acting like my grandson because he is controlling it."

These comments do not show that users would literally mistake a robot for a relative. They show how control provenance changes the meaning of an action. A paw movement generated at random is a robot behavior. The same movement intentionally sent by a son or grandchild can become a personal gesture. The pet's body gives that gesture a place, duration, and tactile possibility.

P2 contrasted a petlike wake-up interaction with a smartphone alarm: "I prefer my smartphone to be woken up, but a little pet that could snooze into my face would be much more sympathetic … have a little paw on you. That would be wonderful." The example is mundane, but it clarifies the design space. Embodiment can make an action feel gentler or more companionable. It also introduces risk: a moving device close to a person's face or body must be predictable, easy to stop, and physically safe.

P3 described a related sense of reassurance: "You feel more secure if you have something … You take it in your arms, and you look at it and hug it, giving you a different feeling as if you're talking to a phone." Holding and hugging change the interaction from attending to a distant interface into engaging with an object in the local environment. This embodied quality is the strongest reason to explore a pet form rather than simply add another video-calling device.

## 4.3 Participants combined communication, companionship, and assistance

Participants did not define one stable role for the robot. They moved among three roles: a communication medium for other people, a companion in its own right, and a device that performs practical tasks. P3 wanted conversation, family communication, messaging, and wake-up support. P4 imagined help with walking or sending a message. P5 wanted companionship and conversation when alone as well as help with physical tasks. P2 extended the idea to demanding household work, including taking out rubbish or clearing snow.

Some expectations exceed what a small, soft robotic pet could safely or economically do. Their importance is analytical rather than contractual. They show that participants evaluated the concept through everyday needs, not through a narrow technical category. A design presented as a "robotic pet" may evoke expectations associated with animals, voice assistants, telepresence systems, mobility aids, and domestic robots at the same time.

P6 expressed this breadth directly: "It could be all. You have to have that in all possibilities." No single device can satisfy every possibility. A more useful interpretation is that the platform should support appropriation. Core communication functions could be combined with optional modules, routines, and settings that fit a particular person and household. Designers should also state the system's limits clearly so that an affectionate form does not imply capabilities it lacks.

## 4.4 Acceptance depended on practical fit and human control

Participants raised requirements that concerned the robot as an object in the home. Surfaces should be easy to clean; components that touch the user should support hygienic maintenance. Voice, movement, and responsiveness should be adjustable. Some participants wanted the pet to follow its owner, which could provide companionship or reassurance but would require reliable navigation and a clear way to pause following.

Voice was especially important because it shaped the pet's perceived character. Participants wanted interaction to be intuitive for people unfamiliar with complex technology. P6 noted that some older adults might need to learn how to interact with the pet before it became useful. This need does not support a one-time "simple interface" solution. It points to onboarding, repair, reminders, and support over time.

The remote-control scenarios also create a consent problem that participants' positive descriptions do not resolve. A family member's ability to move or speak through an object in another person's home can express care, but it can also enable monitoring or intrusion. The local user needs visible and audible cues that identify who is connected and whether an action is remotely generated. They also need convenient controls to accept, end, mute, schedule, or restrict access. The remote partner needs feedback about whether the recipient is available and whether a gesture was delivered. These requirements follow from the relational character of the device: both people participate, and neither should have invisible or unilateral control.

# 5 DESIGN IMPLICATIONS

## 5.1 Design the relationship endpoint as well as the robot

A robotic pet for mediated connection is at least a two-sided system. The local experience includes the pet, its behavior, its place in the home, and controls for receiving contact. The remote experience includes selecting a person, composing a message or gesture, understanding the robot's current state, and knowing whether contact is welcome. Designing only the pet leaves half of the interaction unspecified.

The remote interface should make the social meaning of actions explicit. A sender might choose among a spoken message, a call request, a brief movement, or a touch-like gesture. The interface should preview what the robot will do and state whether the action is immediate or requires acceptance. The local pet should identify the sender before performing the action. This preserves the feature participants valued: an action feels personal because its human source is known.

## 5.2 Preserve reciprocal agency and consent

Remote embodiment changes access to the home. The local user should control when the robot can receive calls, move, use a camera or microphone, and enter particular rooms. Physical controls should remain available even when voice interaction or an app fails. A prominent pause or privacy control can stop movement and communication without requiring the user to navigate a menu.

Reciprocity also means that the older adult should be able to initiate contact through the pet. Designs that mainly allow relatives or caregivers to "check in" risk turning relationship support into surveillance. Contact lists, spoken commands, or a small number of tangible controls could let the local user send a message, request a call, or transmit a gesture. Work on mediator robots and user-led HRI similarly shows that communication depends on giving older adults meaningful control over what is sent and how it is conveyed [26, 32].

## 5.3 Make remote action and automation distinguishable

A hybrid robotic pet may combine direct remote control, prerecorded messages, scheduled routines, and autonomous behavior. Users should be able to tell these sources apart. The distinction could be communicated through a stable combination of voice, light, sound, or movement. For example, the pet might announce "Sara sent you a nudge" before a brief gesture, while autonomous reminders use a different cue.

Legibility supports both emotional meaning and safety. If a user knows that a family member initiated a gesture, the gesture can carry that relationship. If the source is unclear, the same movement may feel confusing or manipulative. Logs should be limited and understandable: users may want to review missed messages, but continuous records of activity in the home would introduce disproportionate privacy risks.

## 5.4 Support customization without requiring constant configuration

Participants differed in desired voice, movement, tasks, and degree of companionship. The pet should allow adjustment of volume, speaking rate, movement intensity, follow behavior, touch response, and notification timing. These choices should be available without making setup a permanent burden. A small set of tested profiles, followed by gradual adaptation, may be more usable than a large settings panel.

Physical design also requires attention. A touchable pet needs removable or cleanable surfaces, stable posture, quiet actuation, safe forces, and behavior that remains predictable around mobility aids and furniture. The design should communicate capabilities honestly. A small companion device may support messaging and reminders; it should not imply that it can lift, carry, or respond to emergencies unless those functions have been validated.

## 5.5 Evaluate relationships over time

Future studies should evaluate the complete relationship system with older adults and their chosen communication partners. Short demonstrations can capture first impressions but cannot show whether the pet finds a durable place in everyday routines. A longitudinal deployment should examine initiation patterns, missed or refused contacts, privacy adjustments, breakdowns, maintenance work, and changes in meaning as novelty fades.

Evaluation should separate proximal interaction outcomes from health outcomes. Relevant near-term measures include ease of initiating contact, perceived social presence, enjoyment, burden, autonomy, and the quality of specific encounters. Claims about loneliness, depression, or well-being require appropriate validated measures, comparison conditions, and sufficient exposure. Table 2 summarizes the connection between the interview evidence and these design directions.

**Table 2: From interview evidence to design requirements.**

| Interview evidence | Design requirement | Evaluation question |
|---|---|---|
| Participants combined phone, messaging, and video according to purpose | Integrate with existing channels and contacts | Does the pet add value without duplicating or complicating current practices? |
| A gesture mattered because a relative controlled it | Identify the human source of every remote action | Can users tell who initiated an action and what it will do? |
| Holding, following, and a paw gesture distinguished the pet from a phone | Use embodiment selectively for presence, proximity, and touch-like expression | Do embodied actions increase perceived presence without causing discomfort or risk? |
| Participants wanted both companionship and communication | Allow roles and functions to change by person and context | Can users appropriate the pet without misunderstanding its capabilities? |
| Participants requested adjustable voice, movement, and easy cleaning | Provide accessible settings and maintainable materials | Can users configure, clean, pause, and recover the device independently? |
| Remote movement and communication enter the private home | Give the local user immediate, granular control | Are access, refusal, privacy, and room boundaries understandable to both parties? |

# 6 DISCUSSION

## 6.1 From companion robot to embodied communication media

The findings support a shift in analytical focus. Companion-robot research often asks whether a person relates to a robot. Our participants also asked how a robot might help them relate to another person. In their scenarios, the robotic pet's meaning depended on a chain of action: a family member sent a voice, movement, or touch-like signal; the robot enacted it locally; and the older adult interpreted the action in light of that relationship.

We call this arrangement embodied communication media. The term emphasizes three properties. First, the device mediates human-to-human communication. Second, its body contributes meaning through movement, location, softness, and touchability. Third, it can still be treated as a petlike social object when no remote person is present. The roles are layered rather than mutually exclusive.

To make the concept usable beyond this study, we propose a four-part sensitizing framework. Relational purpose asks what social act the system is meant to support: a message, a call, a low-intensity gesture, reassurance, shared activity, or companionship. Source of agency asks whether an action is initiated locally, by a remote partner, by automation, or through mixed initiative. Mode of embodiment identifies what the robot adds beyond another screen, including voice, movement, proximity, orientation, and touch-like action. Governance covers consent, provenance, privacy, access, maintenance, and repair. Figure 1 summarizes the framework. It is not a validated taxonomy; it is a compact set of questions for describing prototypes consistently and comparing future studies.

This framing helps explain why pet form may matter. A screen can display a face, but a petlike object can be held, can approach, and can perform a small gesture that persists in the recipient's space. The pet may support low-intensity contact that falls between a message and a scheduled call. A brief nudge from a grandchild, for example, could signal attention without demanding an immediate conversation. Whether such an interaction feels caring or intrusive will depend on timing, consent, relationship history, and the transparency of the system.

## 6.2 Augmentation rather than substitution

Participants situated the concept alongside existing communication and in-person contact. This supports an augmentation principle: the robot should create additional ways to express presence, not stand in for relational work. A device marketed as a solution to loneliness can shift responsibility away from families, communities, and care institutions. The present study provides no evidence that a robot can replace contact or improve mental health.

The more defensible opportunity is narrower. An embodied medium may help people initiate contact, share a small gesture, or make an existing call feel more present. This contribution can still be valuable, but it should be evaluated at the level of actual interactions before broader health claims are made.

## 6.3 Ambiguity is both a resource and a risk

Participants moved fluidly between "the pet," a family member acting through it, and a multifunctional household robot. That ambiguity can support imagination and appropriation. It can also confuse responsibility. If the pet speaks, users need to know whether the words come from a relative, a recorded script, or an AI system. If it moves toward someone, they need to know whether the action is remote or autonomous. If it fails, they need to know who can repair it.

Design should preserve playful interpretation while making control provenance and system capability explicit. This is particularly important for users with changing sensory or cognitive abilities and in care settings where staff members may mediate access. Petlike appearance should not be used to conceal surveillance, overstate intelligence, or secure compliance.

# 7 LIMITATIONS AND RESEARCH AGENDA

This study has five principal limitations. First, the sample was small, self-selected, and technically interested. Five of six participants were men, and five were married. Their expectations may differ from those of people who live alone, have limited technology experience, live in residential care, or experience cognitive or sensory impairment.

Second, the interviews concerned a concept rather than interaction with a prototype. Anticipated use can reveal values and expectations, but it does not predict adoption, learning, breakdowns, or long-term meaning. Statements about touch and remote presence should be treated as design hypotheses.

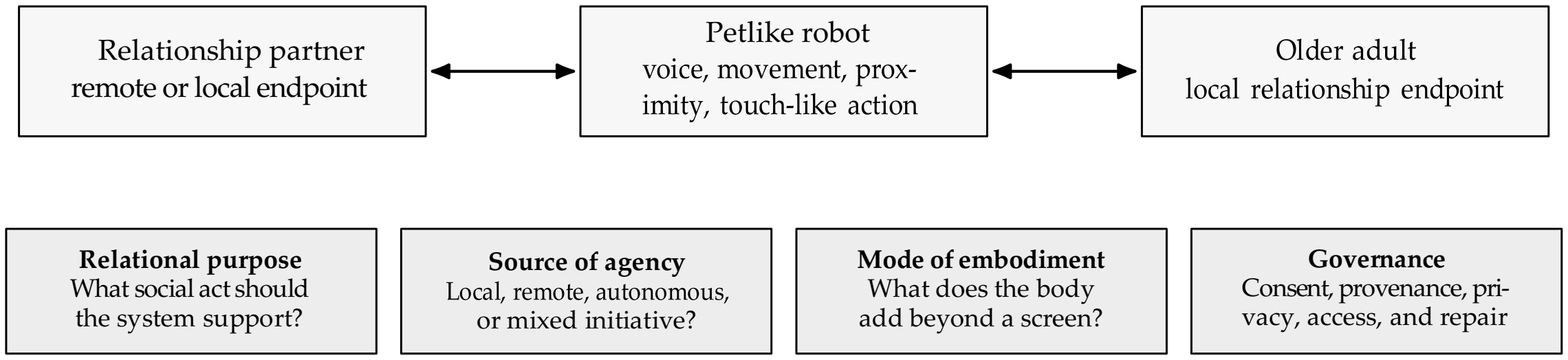

**Figure 1: A sensitizing framework for robotic pets as embodied communication media. Four dimensions characterize a petlike robot used as a relational medium.**

Third, the study included only the prospective local users. Family members, friends, and care workers may have different expectations about effort, availability, privacy, and responsibility. Future work should recruit relationship pairs or networks and examine both sides of mediated interaction.

Fourth, the study did not measure loneliness, depression, relationship quality, or health. Its contribution concerns design possibilities, not clinical or psychosocial effectiveness. Any future outcome study should use validated measures, appropriate comparison conditions, and a duration sufficient to distinguish novelty from sustained value.

Fifth, the published account contains limited detail about analytic roles, translation, and reflexivity. Future studies should report who conducted and coded the interviews, how disagreements were handled, how quotations were translated when applicable, and how researchers' relationships to the project shaped interpretation.

A staged research program follows from these limitations. Early prototypes should test whether users understand remote-control cues, consent controls, and the distinction between human-sent and automated behavior. Dyadic home deployments can then examine appropriation and maintenance over several weeks. Only after usability and relational fit are established should larger studies test effects on social connection or well-being.

# 8 CONCLUSION

Six older adults imagined robotic pets as more than autonomous companions. They described devices that could carry a family member's voice, movement, and touch-like gestures into the home while also offering companionship and practical support. The pet's distinctive value lay in embodiment: it could be approached, held, followed, and associated with a known person in ways that a telephone or smartphone could not fully reproduce.

These findings motivate the concept of robotic pets as embodied communication media. Designing such systems requires attention to both sides of the relationship, reciprocal initiation, visible consent, clear control provenance, customization, hygiene, and realistic capability boundaries. The study does not show that robotic pets improve mental health. It identifies a more precise question for future research: when, for whom, and under what conditions can petlike embodiment make communication between people feel more present without reducing autonomy or privacy?

# ACKNOWLEDGMENTS

We thank the participants for their time and contributions. This work was supported by the German Federal Ministry of Education and Research as part of the BeBeRobot project (grant 16SV8341). The project partners were Osnabrück University, SIBIS Institute for Social Research and Project Consultancy, OFFIS Institute for Information Technology, the German Caritas Association, and the University of Siegen.